# Three-Dimensional Photoacoustic Tomography using Delay Multiply and Sum Beamforming Algorithm

Roya Paridar[1], Moein Mozaffarzadeh[1], Ali Mahloojifar[1], Mohammadreza Nasiriavanaki[2], and Mahdi Orooji[*,1]

[1]Department of Biomedical Engineering, Tarbiat Modares University, Tehran, Iran
[2]Department of Biomedical Engineering, Wayne State University, Detroit, Michigan, USA

## ABSTRACT

Photoacoustic imaging (PAI), is a promising medical imaging technique that provides the high contrast of the optical imaging and the resolution of ultrasound (US) imaging. Among all the methods, Three-dimensional (3D) PAI provides a high resolution and accuracy. One of the most common algorithms for 3D PA image reconstruction is delay-and-sum (DAS). However, the quality of the reconstructed image obtained from this algorithm is not satisfying, having high level of sidelobes and a wide mainlobe. In this paper, delay-multiply-and-sum (DMAS) algorithm is suggested to overcome these limitations in 3D PAI. It is shown that DMAS algorithm is an appropriate reconstruction technique for 3D PAI and the reconstructed images using this algorithm are improved in the terms of the width of mainlobe and sidelobes, compared to DAS. Also, the quantitative results show that DMAS improves signal-to-noise ratio ($SNR$) and full-width-half-maximum ($FWHM$) for about 25 $dB$ and 0.2 $mm$, respectively, compared to DAS.

## 1. INTRODUCTION

Photoacoustic Imaging (PAI), is a promising medical imaging modality that uses an electromagnetic excitation pulse, illuminating the imaging target, to generate photoacoustic (PA) signals.[1,2] It combines the physics of ultrasound (US) and the optical imaging and provides the optical contrast and ultrasonic resolution.[3,4] In addition, it does not have the speckles that appear in a pure US imaging system and is a non-ionizing and noninvasive hybrid imaging modality.[5] PA can be used in different applications, and its effects have been extensively investigated in various cases of studies.[6–10] A numerical analysis for infant brain imaging has been conducted.[11] There are two imaging methods in PAI: PA tomography (PAT) and PA microscopy.[12,13] PAT is a three-dimensional (3D) fast-developing medical imaging technique that is suitable for *in vivo* imaging. Nowadays, PAT is one of the largest research areas in biomedical applications and is still growing quickly.[13,14] As the optical absorption strongly depends on physiological conditions, such as haemoglobin concentration, PAT provides functional information. PAT has been used in several applications such as tumor detection,[15] imaging small animals,[16] blood flow measurement,[17] functional and structural imaging[18] and skin lesions imaging.[19] In PAT, a different types of transducers such as linear, circular and arc, are located around the medium in order to measure the acoustic propagated waves, caused by the laser illumination. Then, a reconstruction algorithm is necessary to form the optical absorption distribution map of the tissue.[20] Low-cost PAI systems using have been designed for different purposes.[21–23] For the linear-array imaging, there is a high similarity between the beamforming algorithm used in PA and US image formation.[24] Delay-and-sum (DAS) beamformer is the most typical algorithm used in medical PA and US imaging, due to its simple implementation. However, the quality of the image obtained by this algorithm is not satisfying, having high level of sidelobes and a low resolution. Delay-multiply-and-sum (DMAS), introduced by Matrone *et al.*,[25] was proposed to improve the image quality compared to DAS. The sidelobes and the width of mainlobe would be reduced using DMAS algorithm. Double stage DMAS (DS-DMAS ) has been developed for linear-array imaging, providing a higher contrast compared to DMAS beamformer.[26–28] In order to improve the resolution of DMAS, it has been combined with Minimum

---

Variance (MV) beamformer.[29–34] As a weighting method for the PA image reconstruction, modified coherence factor (MCF) and high resolution CF (HRCF) have been introduced, which result in a higher contrast and a better resolution compared to the conventional CF, respectively.[35, 36] In this paper, we propose to use DMAS beamforming method for 3D PAI while its superiority for 2D imaging has been proved in the former publications. A 2D array of US transducers is used to perform the 3D PAI. It has been shown that DMAS can be used for 3D application of PAI as well. The rest of the paper is organized as follows. In section 2, a brief explanation about the PA effect and beamformers are presented. In section 3, the results of the numerical study are evaluated. Finally, the conclusion is reported in section 4.

## 2. METHODS

### 2.1 Photoacoustics

The PA waves are generated as follows:

$$\left(\nabla^2 - \frac{1}{c_0^2}\frac{\partial^2}{\partial t^2}\right)p(\vec{r},t) = -\frac{\beta}{C_p}\frac{\partial H(\vec{r},t)}{\partial t}, \quad (1)$$

where $p$ is the initial pressure at position $r$ and time $t$, $C_p$ is the specific heat capacity, $\beta$ is the isobaric volume expansion, $c_0$ is the speed of sound, $H$ represents the heating function which can be interpreted as the product of the spatial absorption function $A(r)$ and the temporal illumination function $I_e(t)$ as below:

$$H(\vec{r},t) = A(r)I_e(t). \quad (2)$$

Assuming that the medium is acoustically homogeneous and $I_e(t) = \delta(t)$, the measured initial pressure can be expressed as:

$$p(\vec{r_0},t_0) = \frac{\beta}{C_p}\int d^3\vec{r}A(\vec{r})\frac{d}{dt_0}\frac{\delta\left(t_0 - \frac{|\vec{r_0}-\vec{r}|}{c_0}\right)}{4\pi|\vec{r_0}-\vec{r}|}. \quad (3)$$

The key problem is about how to reconstruct $A(\vec{r})$ when $p(\vec{r_0},t_0)$ is available. There is a relation between the spherical Radon transform and the initial pressure:

$$g(\vec{r_0},\vec{t}) = \frac{4\pi C_p}{\beta}t\int_0^t dt_0 p(\vec{r_0},t_0) = \int d^3\vec{r}A(\vec{r})\delta(\bar{t} - |\vec{r_0}-\vec{r}|), \quad (4)$$

where $\bar{t} = c_0 t$, and $g(\vec{r_0},\bar{t})$ is the spherical Radon transform of $A(\vec{r})$. The PA image can be reconstructed by inverting (4). There are several methods and algorithms for PA image reconstruction. One of these algorithms is back-projection (BP) algorithm. However, the quality of the reconstructed image by BP in 3D space is not suitable due to artifacts, specially in the case that the number of the active elements in the 2D array sensor is low. In linear-array PAI, BP would be treated as a DAS algorithm.

### 2.2 DAS and DMAS Beamformers

DAS is the most common methods in linear-array medical PA and US image reconstruction. In this algorithm, the delays to the signals received by the array elements are calculated proportional to their distances from the target. Then, the delayed signals are summed up to construct the absorption distribution of the medium. The DAS beamformed signal has the following equation:

$$y_{DAS}(k) = \sum_{i=1}^{N} v_i(k - \Delta_i), \quad (5)$$

where $v_i$ is the detected signal, $\Delta_i$ is the time delay for $i^{th}$ detector, $y_{DAS}(k)$ and $k$ are the output signal and the time index, respectively. As mentioned before, this is the most commonly used algorithm due to its simple implementation. However the quality of the images obtained by this algorithm is not satisfying. The image resolution and contrast obtained from (5) are limited due to the wide mainlobe and large sidelobes of the DAS beamformed signals. In order to overcome these limitations, another beamforming algorithm, named DMAS, is proposed. This algorithm improves the image quality significantly compared to DAS. The DMAS beamformed signal is as follows:

$$y_{DMAS}(t) = \sum_{i=1}^{N-1} \sum_{j=i+1}^{N} s_i(t)s_j(t), \quad (6)$$

where $N$ is the number of receivers, $s_i$ and $s_j$ are the delayed signals received by $i^{th}$ and $j^{th}$ detectors, respectively. As shown in ( 6), in this algorithm the delayed receive signals are combinatorially coupled and multiplied before the summation. The amplitude of each multiplication is squared, and to address this problem, following formula is suggested:

$$y_{DMAS}(t) = \sum_{i=1}^{N-1} \sum_{j=i+1}^{N} sign\left(s_i(t)s_j(t)\right) . \sqrt{|s_i(t)s_j(t)|}. \quad (7)$$

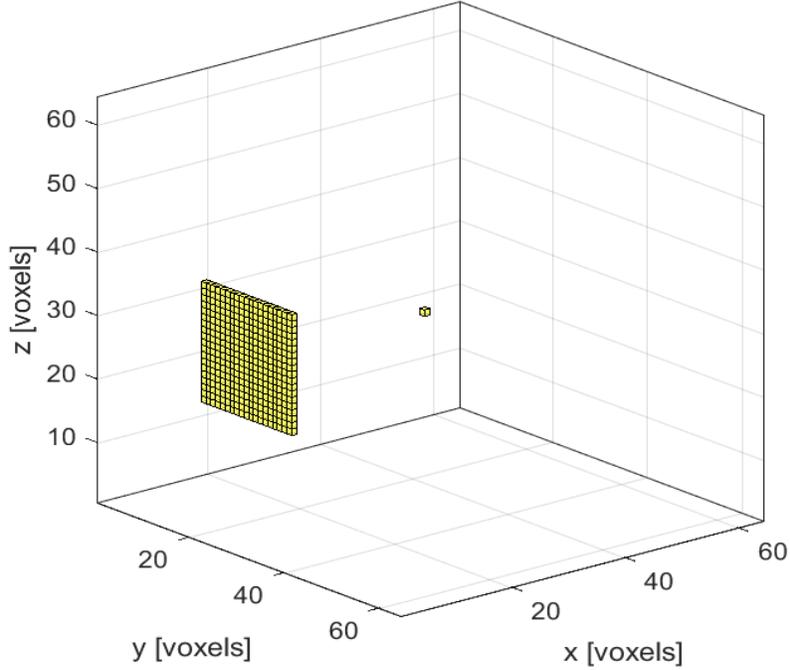

Figure 1: Simulation layout.

Using (7), the sidelobes and the width of mainlobe would be improved compared to (5).

## 3. RESULTS

The k-wave toolbox is used to design the absorber and the 2D array sensor in a 3D space.[37] An imaging region is simulated with dimensions of 6.4 $mm(depth) \times 6.4\ mm(width) \times 6.4\ mm(height)$. A spherical absorber with 0.1 $mm$ radius is located as the initial pressure. The absorber is located at the depth of 3.2 $mm$, and centered on the $y - z$ plane. The speed of sound is assumed 1540 $m/s$. A 2D array of US transducers with 361 elements is located at the center of $y - z$ plane, which both its width and height are 1.9 $mm$. The simulation layout is depicted in Figure 1. Note that the asymmetrically distribution of the array elements with respect to the point target, results in a more noisy reconstructed image; the lateral resolution of the point target will be reduced if its the distance from the center of the array increases .[1] Therefore the array elements are distributed symmetrically with respect to the point target in order to prevent the noise caused by the asymmetric distribution of the elements. The central frequency of the array is 7 $MHZ$ with 77% bandwidth. Gaussian noise is added to the received signals to make the signals similar to the real condition, having a signal-to-noise Ratio ($SNR$) of 50 $dB$ and 10 $dB$. The images are reconstructed using DAS and DMAS algorithms for better comparison.

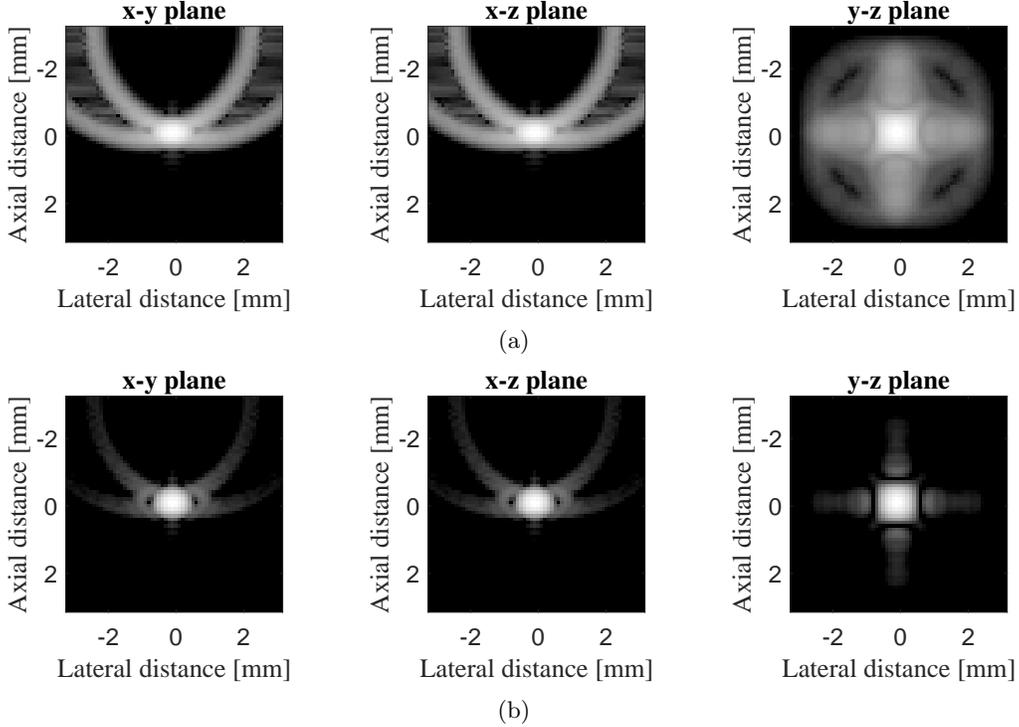

Figure 2: The reconstructed 3D PA images of a single point target using (a) DAS and (b) DMAS algorithm in $x-y$ plane (z=3.2 $mm$), $x-z$ plane (y=3.2 $mm$) and $y-z$ plane (x=3.2 $mm$). Noise is added to the received signals having a $SNR$ of 50 $dB$.

Table 1: $FWHM$ at the depth of 3.2 $mm$

| Beamformer | $FWHM(mm)$ |
|---|---|
| DAS | 0.76 |
| DMAS | 0.55 |

The results are shown in Figure 2 where we face a $SNR$ of 50 $dB$. Figure 2(a) and Fig 2(b) show the reconstructed images using DAS and DMAS beamformer, respectively. The first column shows a cross section of the reconstructed image in the $x-y$ plane, and the second and third column also show the results in the $x-z$ plane and $y-z$ plane, respectively, at the depth of 3.2 $mm$. Note that since the $x-y$ and $x-z$ plane are both perpendicular to $y-z$ plane (where the array sensor is located, named detection plane), the reconstructed images in these two planes are the same. It can be seen that the reconstructed image using DMAS has a higher quality compared to DAS, as expected. Also, a better noise suppression is achieved. To have a better evaluation, lateral variations are presented in Figure 3. Lateral variations at the depth of 3.2 $mm$ in the detection plane and $x-y$ plane (perpendicular to the detection plane) are shown in Figure 3(a) and Figure 3(b), respectively. The lateral variations in $x-z$ plane are not depicted since it is similar to the lateral variations in $x-y$ plane. It is obvious that DMAS beamformer results in a narrower width of mainlobe and lower level of sidelobes compared to DAS, which indicates the improvement in the terms of resolution and contrast. Consequently, DMAS results in an image with a higher quality compared to DAS. $SNR$ and the spatial resolution are two important metrics to evaluate and compare the performance of the beamformers. The full-width-half-maximum ($FWHM$) in -6 $dB$ can be used to estimate the spatial resolution. The $FWHM$ and $SNR$ for the single point target at the depth of 3.2 $mm$ (shown in Figure 2) are calculated and presented in Table 1 and Table 2, respectively. Table 1 shows that DMAS provides a narrower width of mainlobe in -6 $dB$ compared to DAS. Note that the calculated $FWHM$ are similar in all three planes. $SNR$s are calculated as follows:

$$SNR = 20\log_{10} P_{signal}/P_{noise}, \qquad (8)$$

where $P_{signal}$ is the difference between the maximum and minimum intensity of the reconstructed image, and $P_{noise}$ is the standard deviation of the reconstructed image.[27] Table 2 shows that DMAS beamformer results in a higher $SNR$ compared to DAS. Therefore, DMAS outperforms DAS in the terms of noise reduction. It should be noted that the calculated $SNR$ are similar in $x-y$ plane and $x-z$ plane.

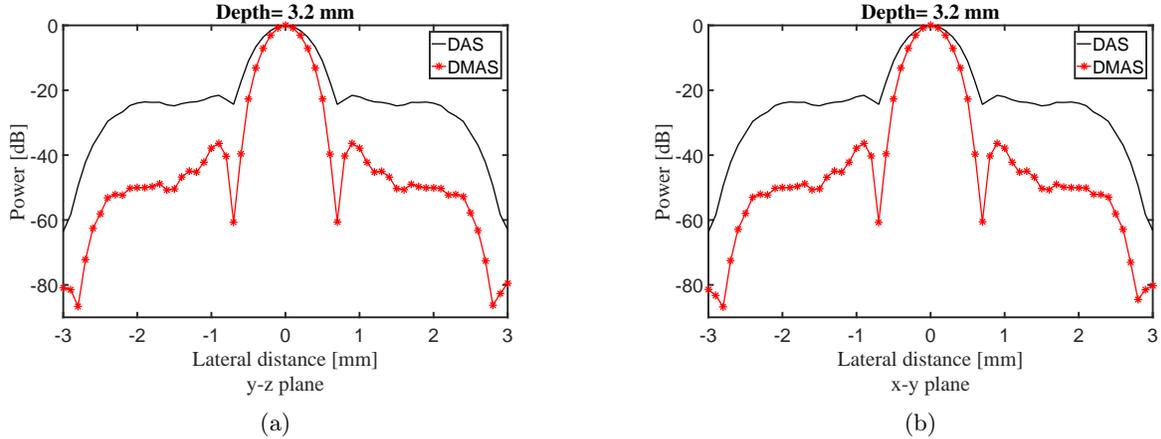

Figure 3: Lateral variations of the reconstructed image of a single point target using DAS and DMAS at the depth of 3.2 $mm$ in (a) $y-z$ plane, named detection plane, and (b) $x-y$ plane which is perpendicular to the detection plane.

Table 2: $SNR$ at the depth of 3.2 $mm$

| Beamformer | $SNR(dB)$ | |
| --- | --- | --- |
| | $y-z$ plane | $x-y$ plane |
| DAS | 38.90 | 42.00 |
| DMAS | 64.14 | 66.63 |

## 4. CONCLUSION

3D PAI is used to make the diagnosis procedure faster and more accurate. In this paper, DAS and DMAS beamforming methods were applied on a 3D imaging space, where the PA signals were acquired by a 2D array of the US transducers. The quality of the reconstructed image obtained by DAS was not satisfying due to its high level of sidelobes and wide mainlobe. It was suggested to use DMAS algorithm in order to improve the image quality compared to DAS. Numerical simulation with a single point target was performed. The reconstructed images were evaluated qualitatively and quantitatively. The qualitative results showed that DMAS algorithm provides an image with a better quality and better noise reduction compared to DAS. Also, quantitative results showed that, compared to DAS, $SNR$ and $FWHM$ were improved for about 25 $dB$ and 0.2 $mm$, respectively.